%% file: engel.tex
\documentclass[12pt]{article}
\usepackage{graphicx}
\usepackage{amsmath,braket,xcolor}

\newcommand\pubnumber{CIPANP2015-Engel}
\newcommand\pubdate{\today}

\def\unc{Department of Physics and Astronomy\\
CB 3255, University of North Carolina, Chapel Hill, NC 27599-3255 USA}
\def\support{\footnote{Work supported 
in part by the U.S.\ Department of Energy, Office of
Science, Office of Nuclear Physics, under Contract Nos.\  DE-FG02-97ER41019 and
DE-SC0008641, ER41896 (NUCLEI SciDAC collaboration).}}

\def\Title#1{\begin{center} {\Large #1 } \end{center}}
\def\Author#1{\begin{center}{ \sc #1} \end{center}}
\def\Address#1{\begin{center}{ \it #1} \end{center}}

\newcommand\pubblock{\rightline{\begin{tabular}{l} \pubnumber\\
         \pubdate  \end{tabular}}}
\newenvironment{Abstract}{\begin{quotation}  }{\end{quotation}}
\newenvironment{Presented}{\begin{quotation} \begin{center} 
             PRESENTED AT\end{center}\bigskip 
      \begin{center}\begin{large}}{\end{large}\end{center} \end{quotation}}
\def\Acknowledgements{\bigskip  \bigskip \begin{center} \begin{large}
             \bf ACKNOWLEDGEMENTS \end{large}\end{center}}

\input econfmacros.tex
\newcommand{\bbz}{$0\nu\beta\beta$}
\newcommand{\bbt}{$2\nu\beta\beta$}
\newcommand{\bb}{$\beta\beta$}

\begin{document}
\begin{titlepage}
\pubblock

\vfill \Title{Nuclear Matrix Elements for Double-$\beta$ Decay} \vfill
\Author{ Jonathan Engel\support} \Address{\unc} \vfill
\begin{Abstract}
Recent progress in nuclear-structure theory has been dramatic.  I describe
recent and future applications of ab initio calculations and the generator
coordinate method to double-beta decay.  I also briefly discuss the old and
vexing problem of the renormalization of the weak nuclear axial-vector coupling
constant ``in medium'' and plans to resolve it.
\end{Abstract}
\vfill
\begin{Presented}
Twelfth Conference on the Intersections of Particle and Nuclear Physics\\
Vail, Colorado, May 19–24, 2015
\end{Presented}
\vfill
\end{titlepage}
\def\thefootnote{\fnsymbol{footnote}}
\setcounter{footnote}{0}

Neutrinoless double-beta ($0\nu\beta\beta$) decay occurs if neutrinos are
Majorana particles, at a rate that depends on a weighted average of neutrino
masses (see Refs.\ \cite{avi08,Ver12} for reviews).  New experiments to search
for $0\nu\beta\beta$ decay are planned or underway.  Extracting a mass from the
results, however, or setting a reliable upper limit, will require accurate
values of the nuclear matrix elements governing the decay.  These cannot be
measured and so must be calculated.

The matrix elements have been computed in venerable and sophisticated models,
but vary by factors of two or three.  All the models can be improved, however.
Here I focus on two of them: the shell model and the generator coordinate method
(GCM).  I first discuss effective interactions and decay operators for the shell
model that will connect that method to ab initio nuclear-structure calculations,
which have made rapid progress recently. I then show how the GCM avoids problems
of the quasiparticle random phase approximation (QRPA) and, moreover, can be
extended so that it incorporates the QRPA's ability to capture proton-neutron
pairing.  Finally, I briefly examine the currently unsettling
``renormalization'' of the nuclear weak axial coupling constant $g_A$, and argue
that the cause will be identified soon through and investigation of many-body
currents and the effective enlargement of model spaces.  

The lifetime for $0\nu\beta\beta$ decay, if the exchange of the familiar light
neutrinos is responsible, is given by the product of a phase space factor
(recently recomputed in Ref.\ \cite{Kot12}) an effective mass $m_\nu = \sum_i
U_{ei}^2 m_i$, where $m_i$ is the mass of the $i^{\rm th}$ eigenstate and
$U_{ei}$ weights each mass by the mixing angle of the associated eigenstate with
the electron neutrino, and $M^{0\nu}$ is the nuclear matrix element.  The matrix
element is complicated but can be simplified without significantly altering its
value through the ``closure approximation.'' In this approximation, and
neglecting two-body currents (which I take up briefly later), one can write the
matrix element as
\begin{align}
\label{eq:me}
M^{0\nu}  &=  \frac{2R}{\pi g_A^2} \int_0^\infty \!\!\! q \, dq \\ 
 \times &\bra{f}  \sum_{a,b}\frac{j_0(qr_{ab})\left[ h_F(q)+   
 h_{GT}(q) \vec{\sigma}_a \cdot \vec{\sigma}_b\right] + 3 j_2(qr_{ab})
 h_T(q) \vec{\sigma}_a \cdot \vec{r}_{ab} \vec{\sigma}_b \cdot \vec{r}_{ab} }
 {q+\overline{E}-(E_i+E_f)/2} \tau^+_a \tau^+_b \ket{i} \,, \nonumber  
\end{align}
where $r_{ab} \equiv |\vec{r}_a-\vec{r}_b|$ is the distance between nucleons $a$
and $b$, $j_0$ and $j_2$ are the usual spherical Bessel functions, $\bar{E}$ is
an average excitation energy to which the matrix element is insensitive, and the
nuclear radius $R\equiv 1.2 A^{1/3}$ fm is inserted with a compensating factor
in the phase-space function to make the matrix element dimensionless.  The
``form factors'' $h_F$, $h_{GT}$, and $h_T$ are given in Refs.  \cite{sim99} and
\cite{sim08}.

As already indicated, researchers have applied a variety of nuclear models to
\bb\ decay.  At the moment, it is difficult to assess the uncertainty in any one
of the matrix-element calculations.  Quantifying uncertainty is an important
task for the next few years, but just as important is reducing the uncertainty
by improving the calculations.  We can do both by linking model Hamiltonians and
decay operators to data through ab initio calculations.  The lighter \bb\ nuclei
($^{76}$Ge, $^{82}$Se), or those such as $^{136}$Xe that are near closed shells,
will be the easiest to connect to ab initio work.  The heavier nuclei will
generally require a different treatment.  I discuss those suitable for an ab
initio treatment first.

The shell model is a complete diagonalization in a subspace of the full
many-body Hilbert space that consists of all possible configurations of valence
particles within a few valence single-particle orbitals, outside a core that is
forced to remain inert.  The model thus neglects excitations of the particles in
the core into the valence levels or higher-lying levels, as well as excitations
of the valence nucleons into higher-lying levels.  At present, practitioners
usually deal with this problem by constructing a phenomenological Hamiltonian
for use in the shell model space.  Nuclear-structure theory is reaching the
point where we can do better, however.  A variety of many-body methods now yield
accurate solutions to the Schr\"{o}dinger equation for nuclei with closed shells
and nuclei with one or two nucleons outside closed shells.  Among the methods
are the coupled clusters approach \cite{jansen11} and the in-medium similarity
renormalization group (IMSRG) \cite{tsukiyama12}.  These approaches can be used
to construct, for up to two or three nucleons in the valence shell, a unitary
transformation that transforms the full many-body Hamiltonian into block
diagonal form, with a piece $H_\text{eff}$ in the shell-model space that
reproduces the lowest-lying energies exactly.  

\begin{figure}[t]
\centering
\includegraphics[width=.95\textwidth]{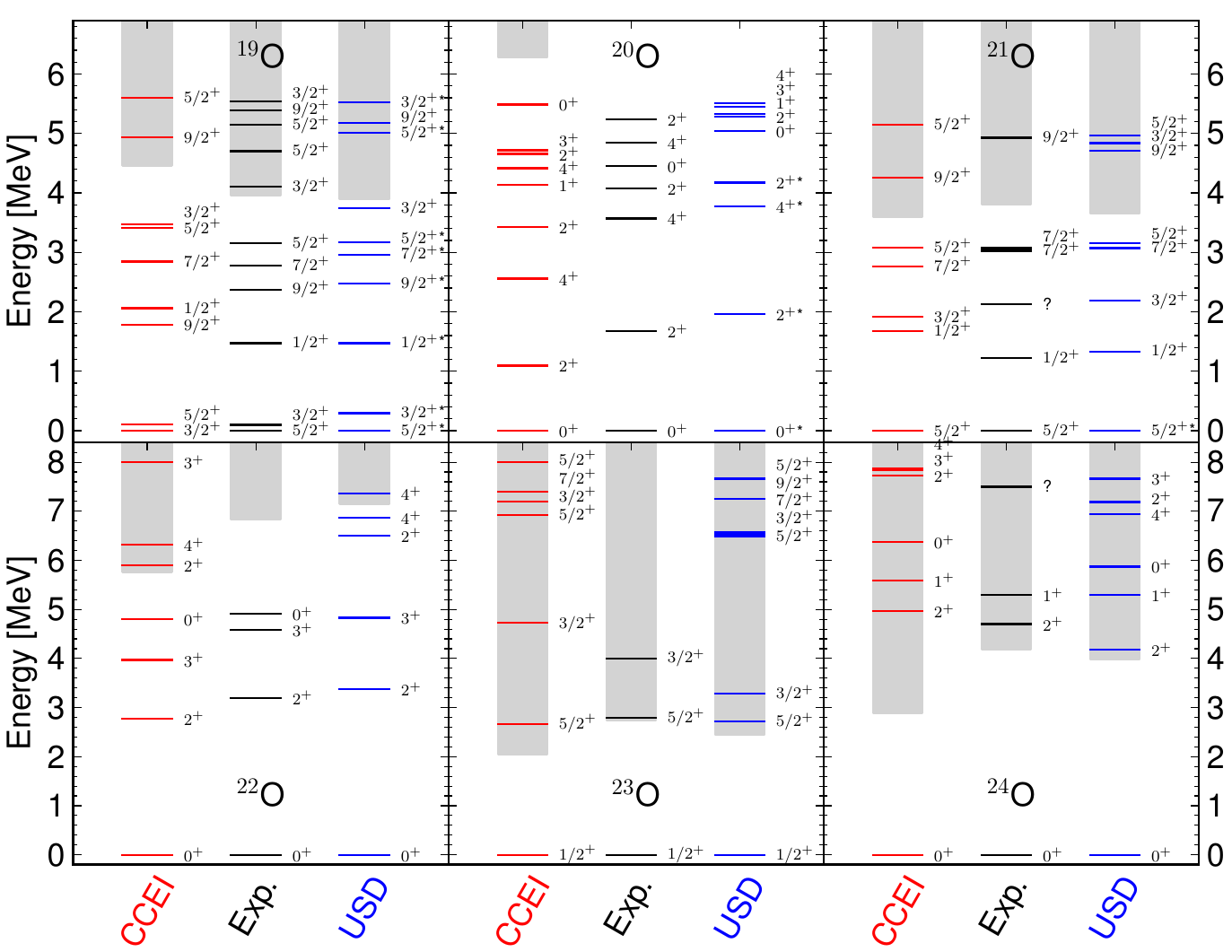}
  \caption{\label{fig:ox}Spectra of neutron-rich oxygen isotopes.  The left
  column contains the results of $sd$-shell-model calculations with an
  effective Hamiltonian derived form ab initio chiral two- and three-body
  forces \cite{n3lo} and coupled-cluster calculations in $^{16,17,18}$O. The
  middle column contains experimental data and the right column contains the
  predictions with the phenomenological USD interaction \cite{bro88} that
  was fit to data in the same shell.}
  \end{figure}

The procedure for obtaining the \bbz\ matrix element in $^{76}$Ge via a
coupled-clusters-based shell-model calculation would go something like this:
\begin{enumerate}
\item Derive a two- and three-nucleon Hamiltonian from chiral effective field
theory \cite{n3lo} or phenomenology in few-nucleon systems. 
\item Do ab initio coupled-clusters calculations of the ground state of the
closed shell nucleus $^{56}$Ni, of the low-lying eigenstates states of the
closed-shell-plus-one nuclei $^{57}$Ni and $^{57}$Cu, and of the low-lying
states of the closed-shell-plus-two nuclei $^{58}$Ni, $^{58}$Cu, and $^{58}$Zn.
Eventually, when it becomes possible, do the same in closed-shell+three nuclei
as well.
\item Perform a ``Lee-Suzuki'' mapping \cite{lis09} of the low-lying states in
these nuclei onto states in the valence shell containing one and two (and
eventually, three) nucleons.  The mapping is designed to maximize the overlap of
the full ab initio eigenstates with their shell-model images, while preserving
orthogonality of the images \cite{kvaal}.
\item Use the mapping of states to construct the shell-model interaction $H_{\rm
eff}$ that gives the image states the same energies as their parents.  Construct
an effective double-beta operator that gives the same matrix elements between
image states as the bare operator does between the associated parents.
\item Put 4 protons and 16 neutrons (for $^{76}$Ge) and 6 protons and 14
neutrons (for $^{76}$Se) in the valence shell and use the effective interaction
and decay operator derived in the previous step to calculate the
ground-state-to-ground-state decay matrix element.
\end{enumerate}

We have just begun to carry out this program \cite{jansen14}, starting in
lighter nuclei.  Using coupled cluster calculations in $^{16,17,18}$O, we
predicted the spectra of oxygen isotopes with more neutrons.  Fig.  \ref{fig:ox}
shows the results.  The left column for each isotope contains our predictions,
the middle column the experimental data, and the right column the
``predictions'' of the USD shell-model interaction \cite{bro88} that was
fit long ago to lots of data \emph{in the sd shell itself}.  Our interaction,
which uses only data in two- and three-nucleon systems produces results which
are at least as good.  Though we have yet to investigate matrix elements of the
double-beta operator, these initial results for energy levels are extremely
promising.  Including an effective three-nucleon interaction, from still
extremely difficult ab initio calculations in $^{19}$O and $^{17}$C (ensuring
that our predictions exactly match the ab initio results in those isotopes)
should improve the spectra further.  A larger shell model space would do the
same.  A similar program is being undertaken within the IMSRG \cite{bog14}.  

\begin{figure}[t]
\centering
\includegraphics[width=.7\textwidth]{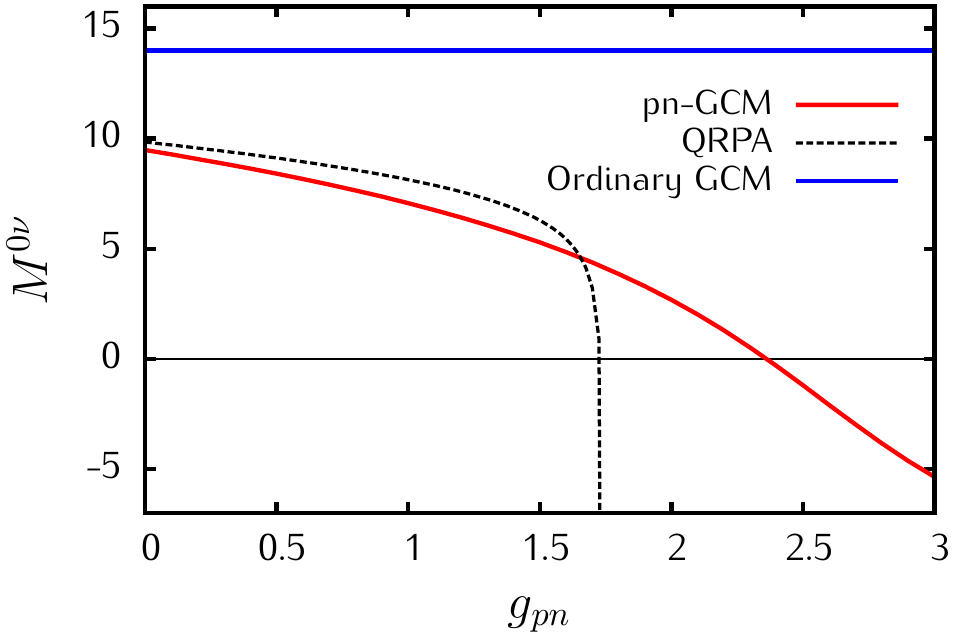}
\caption{\label{fig:gcmqrpa} \bbz\ matrix element for decay of $^{76}$Ge in
three models: the pn-QRPA (black dashed curve), the ordinary GCM without
explicit proton-neutron correlations (flat blue solid line), and the GCM with
proton-neutron pairing correlations (solid red curve).  The quantity $g_{pn}$
is the strength of the proton-neutron pairing in the Hamiltonian in Eq.\
\eqref{eq:h}.  The quadrupole interaction in that Hamiltonian is turned off for
this illustration.  The calculations use two full oscillator shells as the
single-particle model space.}
\end{figure}

For heavier complicated nuclei such as $^{130}$Te or $^{150}$Nd, fully ab initio
calculations in the region are still a ways off.  In those, we may have to use a
more restricted wave function.  Fortunately, recent work suggest that collective
correlations may be most of what you need for an accurate matrix element
\cite{menendez15}.  The phenomenological methods, e.g., density functional
theory, are built for collective correlations.  

Such methods, particularly the GCM, have already been applied to \bb\ decay
\cite{rod10,rod11}, but not all collective correlations have been included.
Neutron-proton pairing, in particular, is omitted because its effects are hard
to see in nuclear spectra and transitions.  It does, however, play a significant
role in \bb\ decay.  To see this, we have carried out calculation of the decay
of $^{76}$Ge in a Hilbert space consisting of 36 nucleons in two full oscillator
shells with a semi-realistic interaction of the form
\begin{equation}
\label{eq:h}
H =h_0 -\sum_{\mu=-1}^1  g^{T=1}_\mu \ S^\dag_\mu S_\mu -\frac{\chi}{2}
\sum_{K=-2}^2 Q^\dag_{2K} Q_{2K} 
- g_{pn} \sum_{\nu=-1}^1 P^\dag_\nu P_\nu + g_{ph} \sum_{\mu,\nu=-1}^1
F^{\mu\dag}_\nu F^\mu_\nu \,,
\end{equation}
where $h_0$ contains single particle energies, the $Q_{2K}$ are components of
the quadrupole operator, and
\begin{equation}
\label{eq:h-ops}
S^\dag_\mu  = \frac{1}{\sqrt{2}} \sum_l \sqrt{2l+1} [c^\dag_l
c^\dag_l]^{001}_{00\mu} \,, \quad P^\dag_\mu = \frac{1}{\sqrt{2}} \sum_l 
\sqrt{2l+1} [c^\dag_l c^\dag_l]^{010}_{0\mu 0} \,, \quad
F^{\mu}_{\nu} = \frac{1}{2} \sum_{i}\sigma^{\mu}(i) \tau^{\nu}(i) \,. 
\end{equation}
In this last line $c^\dag_l$ is a creation operator, $l$ labels single-particle
multiplets with good orbital angular momentum, $S^\dag_\mu$ creates a correlated
pair with total orbital angular momentum $L=0$, spin $S=0$, and isospin $T=1$
(with $\mu$ labeling the isospin component $\tau=T_z$), $P^\dag_\mu$ creates an
isoscalar proton-neutron pair with $L=0$ and $S=1$ ($S_z=\mu$), and the
$F^\mu_\nu$ are the components of the Gamow-Teller operator.  The Hamiltonian
incorporates like-particle and proton-neutron pairing, a quadrupole-quadrupole
interaction, and a repulsive ``spin-isospin'' interaction.  Ref.\
\cite{menendez15} shows that in the $fp$ shell, anyway, this kind of interaction
reproduces full shell-model results accurately.  
 
Fig.\ \ref{fig:gcmqrpa} shows the GCM results for the decay of $^{76}$Ge; the
quadrupole interaction is temporarily turned off.  The dashed curve is from the
QRPA; it blows up when the proton-neutron pairing strength is near 1.5 (a
realistic value).  The reason is that the mean field on which the QRPA is based
undergoes a phase transition from a condensate of like-particle pairs to a
condensate of proton-neutron pairs at that point.  The QRPA is unable to
accommodate more than one mean field; it breaks down at the transition point.
The GCM on the other hand, is explicitly designed to mix many mean fields.  The
technique is usually applied to nuclei that don't have a definite shape (many
\bb-decay candidates are in this class), with wave functions that are
superpositions of states with a range of deformation.  The GCM generates mean
fields with that range by minimizing the mean-field energy under the constraint
that the quadrupole moment take a particular value, then repeating the
minimization for lots of other values for the quadrupole moment.  The
interaction is then diagonalized in the space of constrained mean-fields,
usually after each has been projected onto states with well defined angular
momentum and particle number.

The method as just described was applied together with the phenomenological
density-dependent Gogny interaction to \bbz\ decay in Refs.\ \cite{rod10,rod11}.
The resulting matrix elements are usually larger than those of the shell model
or QRPA.  One reason is the absence of the proton-neutron correlations that in
the QRPA shrink the matrix element as in Fig.\ \ref{fig:gcmqrpa} (before ruining
it completely for very strong proton-neutron pairing).  But one can add the
physics of proton-neutron pairing by mixing together mean-fields (quasiparticle
vacua) with different degrees of that pairing.  One does so by imposing
constraints on both the quadrupole moment and the proton-neutron pairing
amplitude, that is by minimizing 
\begin{equation}
\label{eq:constrained-hfb}
H'=H - \lambda_Z N_Z - \lambda_N N_N - \lambda_Q Q_{20} -
\frac{\lambda_P}{2} \left( P_0 + P_0^\dag \right) \,,
\end{equation}
where the Lagrange multipliers $\lambda_Z$ and $\lambda_N$ fix the expectation
values of the proton and neutron number operators $N_Z$ and $N_N$ --- this is
part of the usual HFB minimization --- and the other multipliers fix the
quadrupole moment $\braket{Q_{20}}$ and the proton-neutron pairing amplitude
$\phi\equiv\braket{P_0+P_0^\dag}$.  Such a minimization requires generalizing
the usual BCS-like wave functions to include proton-neutron, leading to
quasiparticles that are part proton and part neutron as well part particle and
part hole (which they are even in the usual treatment).

\begin{figure}[t]
\includegraphics[width=.485\textwidth]{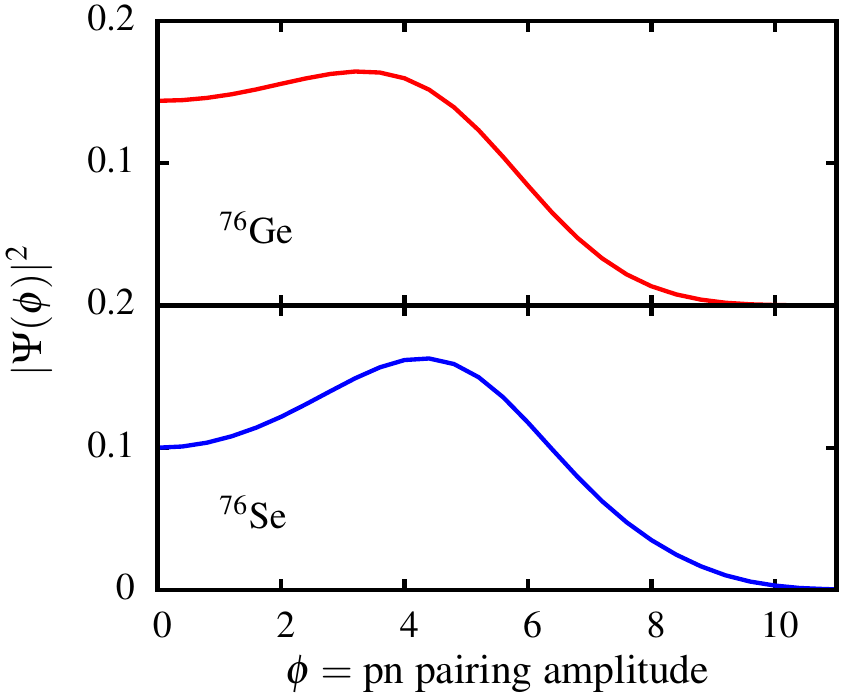}%
\hspace{.03\textwidth}\includegraphics[width=.485\textwidth]{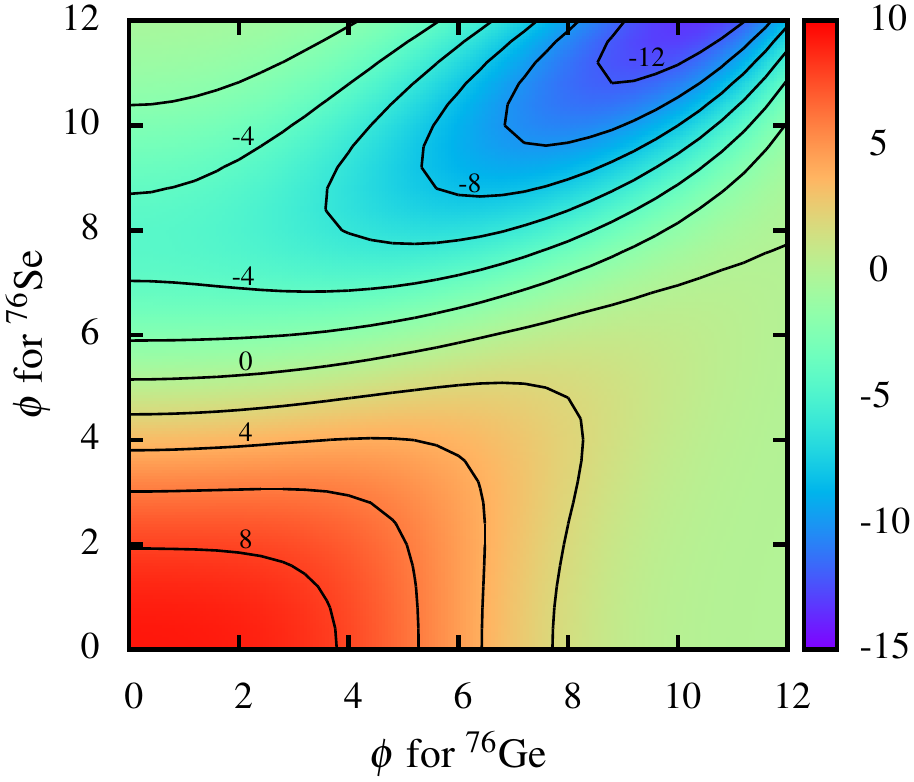}
\caption{\label{fig:wfs}Left: Squares of collective wave functions, as a
function of the proton-neutron pairing amplitude $\phi$ in $^{76}$Ge (top) and
$^{76}$Se (bottom), the initial and final nuclei in the decay of $^{76}$Ge, for
the same Hamiltonian \eqref{eq:h} as used for Fig.\ \ref{fig:gcmqrpa}.  Right:
\bbz\ matrix element as a function of the pairing amplitudes in the projected
mean-field states making up the ground states of the initial and final nuclei.
The matrix element has a maximum for no proton-neutron pairing and is reduced
at the point at which the collective wave functions peak.}
\end{figure}

Ref.\ \cite{hin14} carries out this calculation.  With the quadrupole moment
turned off, it produces ``collective wave functions'' of the proton-neutron
pairing amplitude, the squares of which appear on the left side of Fig.\
\ref{fig:wfs}.  These represent the probability that the final diagonalized
ground states in Ge and Se contain a given proton-neutron pairing amplitude
$\phi = \braket{P_0+P_0^\dag}$.  One can see that the wave functions are peaked
around $\phi=4$ or 5.  The right side of the figure shows the \bbz\ matrix
element as a function of the two pairing amplitudes.  At the point representing
the peak of the two wave functions, the matrix element is noticeably smaller
than the point at which the pairing amplitudes are zero.  Finally, the part of
Fig.\ \ref{fig:gcmqrpa} I haven't focused on shows the \bbz\ matrix element as a
function of the proton-neutron pairing strength.  The GCM curve mirrors that
produced by the QRPA until a point close to the mean-field phase transition,
around and after which it behaves smoothly (as it should; there's no real phase
transition beyond mean-field theory).  The matrix element is indeed smaller than
that of the ordinary GCM, which captures no proton-neutron correlations of this
type and thus produces a result that is independent of $g_{pn}$.

The next step in the development of this approach is to move beyond the
semi-realistic calculation discussed here and marry this enlarged GCM with
sophisticated Skyrme or Gogny density functionals, which work in complete
single-particle spaces, with all the nucleons active.  The result will almost
certainly be matrix elements that are closer to those of the shell model.

I turn finally to the renormalization of the axial-vector coupling $g_A$.  It
has been know for some time (see, e.g., Ref.\ \cite{bro88}) that matrix elements
for $\beta$ and \bbt decay are smaller in reality than in our calculations.  If
\bbz\ matrix elements are as small compared to our calculations as \bbt matrix
elements, experiments are in trouble.  Fortunately, the issue can now be
investigated systematically.  There can only be two sources of the quenching:
many-body weak currents, which would alter the predictions of calculations with
the one-body Gamow-Teller operator, and model space truncation, i.e.\ the
omission of important configurations.  Work is now beginning to examine both
these sources.  The effects of many-body currents have traditionally been
thought to be small \cite{tow87}, but the construction of those currents in
chiral effective field theory --- currents that should go along with the
interactions used by ab initio calculations --- may lead to larger effects
\cite{men11,eng14}. Crucially, however, those effects should be smaller for
\bbz\ decay than for \bbt\ decay.  The issue should be cleared up by careful EFT
parameter fits in the near future.  The other source of quenching, model-space
truncation, can be investigated in the ab initio shell-model calculations
described earlier.  Those implicitly include many configurations from outside
the model space in the effective interactions and operators.  We should soon be
able to see whether \bbz\ decay is quenched, and if so, by how much.  

\Acknowledgements I am grateful to my collaborators G.R.\ Jansen, G. Hagen, N.\
Hinohara, M.\ Mustonen, P.\ Navratil, A.\ Signoracci, F. \v{S}imkovic, and P.\
Vogel.  This work was supported in part by the U.S.\ Department of Energy,



\end{document}

%% file: econfmacros.tex
\def\beq{\begin{equation}}
\def\eeq#1{\label{#1}\end{equation}}
\def\eeqn{\end{equation}}

\def\beqa{\begin{eqnarray}}
\def\eeqa#1{\label{#1}\end{eqnarray}}
\def\eeqan{\end{eqnarray}}

\let\bar=\overbar

\def\bra#1{\left\langle{ #1} \right|}
\def\ket#1{\left| {#1} \right\rangle}

\def\Dslash{\not{\hbox{\kern-4pt $D$}}}
\def\dslash{\not{\hbox{\kern-2pt $\del$}}}

\def\msb{{\bar{\ssstyle M \kern -1pt S}}}

